\documentclass[a4paper,11pt]{article}
\usepackage{jinstpub} 

\title{\boldmath Characterization of Residual Charge Images in LSST Camera e2v CCDs}

\author{Daniel Polin,}
\author{Adam Snyder,}
\author{Craig Lage,}
\author{J. Anthony Tyson}
\affiliation{Department of Physics and Astronomy, University of California, Davis, CA 95616}

\emailAdd{dapolin@ucdavis.edu}

\abstract{LSST Camera CCDs produced by the manufacturer e2v exhibit strong and novel residual charge images when exposed to bright sources. These manifest in images following bright exposures both in the same pixel areas as the bright source, and in the pixels trailing between the source and the serial register. Both of these pose systematic challenges to the Rubin Observatory Legacy Survey of Space and Time instrument signature removal. The latter trail region is especially impactful as it affects a much larger pixel area in a less well defined position. In our study of this effect at UC Davis, we imaged bright spots to characterize these residual charge effects. We find a strong dependence of the residual charge on the parallel clocking scheme, including the relative levels of the clocking voltages, and the timing of gate phase transition during the parallel transfer. Our study points to independent causes of residual charge in the bright spot region and trail region. We propose potential causes in both regions and suggest methodologies for minimizing residual charge. We consider the trade-offs to these methods including decreasing the camera's full well and dynamic range at the high end. These results and suggestions have been reviewed by the camera commissioning team and may result in changes made to the clocking voltage scheme on the LSST Camera.}


\arxivnumber{2502.05418}

\begin{document}
\maketitle
\flushbottom

\section{INTRODUCTION}
\label{sec:intro}
The NSF-DOE Vera C. Rubin Observatory Legacy Survey of Space and Time (LSST) depends on the ability to resolve astronomical sources at faint signal levels\cite{ivezic2019lsst}. When the system is calibrated to image objects as faint as 24.5 magnitude in a 30s exposure over 9.6 square degrees, we will inevitably image bright objects above the dynamic range of the camera. LSST will regularly include bright stellar images and bright transient objects like satellites\cite{Tyson2020} that exceed the saturation point of the Rubin Observatory LSST Camera (LSSTCam) charge-coupled devices (CCDs). This type of exposure has associated sensor effects that must be characterized and mitigated in the survey for the faint science programs to succeed.

One systematic sensor effect common to CCDs is residual images seen in exposures immediately following a bright or overexposed image. These image ``ghosts" are a fairly well documented phenomena. The mechanism through which this occurs is fairly straightforward: When CCDs are exposed to a bright object, not all of the electrons produced are captured in the pixel charge packet or read out in that image. This charge instead persists within the CCD after the image is read out. On some time scale which is long relative to a single readout, the charge then thermally leaks back into the pixel's potential well and is read out in subsequent exposures as ghosts of the bright objects~\cite{rest2002residual,janesick1992history}.

In a more specific example, under some operating conditions it is possible for charge to reach the silicon-silicon oxide (Si-SiO$_2$) interface at the CCD surface where the charge can become trapped. Historically these have been called residual surface images to differentiate them from residual charge images stored in the bulk silicon. However, since the latter is not seen in LSSTCam, we will refer to them simply as residual images or residual charge images. These have also been variously called ``image persistence"\cite{Doherty2014,waters2020pan}, and the ``release of trapped charge"\cite{Janesick2001} although the mechanism and effect is believed to be the same.

Residual charge images occur when the parallel clocking barrier and collecting voltage levels are configured in such a way that stored charge reaches the surface  before it overflows into neighboring pixels in the process called ``blooming". The higher collecting voltage level (parallel high), if set high enough, pulls electrons towards the surface. The relative level of the barrier voltage (parallel low) then determines the maximum amount of charge that can be stored in a given pixel before it blooms, which we call the ``blooming full well". Increasing the number of electrons in a given pixel will also push charge packets further towards the surface. If the full well is large enough that charge reaches the surface, we call this the ``surface full well condition". Residual charge can then be addressed proactively with changes in the parallel clocking scheme, or after the fact by raising the substrate voltage to inversion which can provide holes at the surface with which these trapped electrons recombine. 

\begin{figure}[h]
    \centering
    \includegraphics[width=0.5\linewidth]{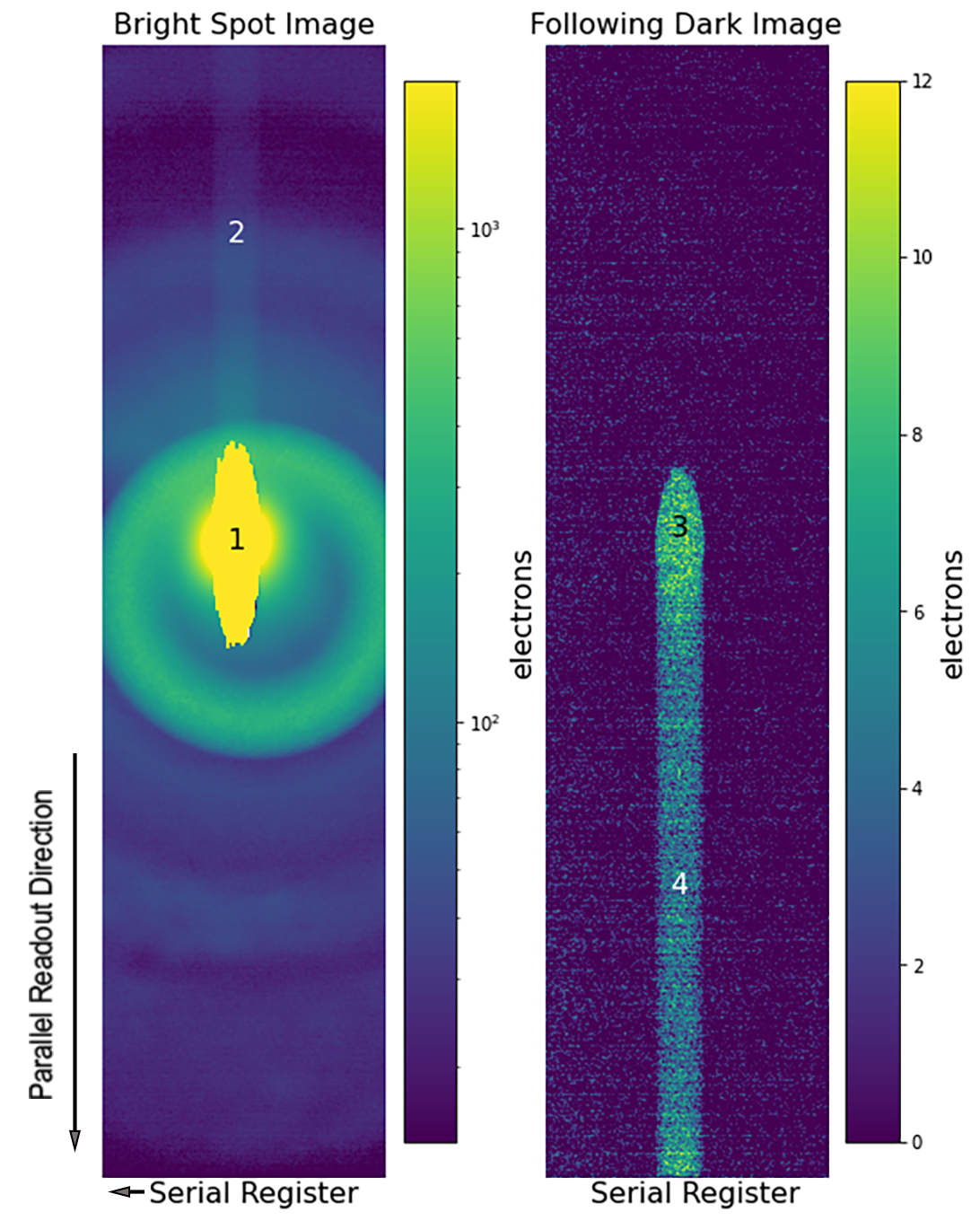}
    \caption{An example of residual charge images seen in LSSTCam e2v CCDs. Two images are shown from the same 2000 $\times$ 512 pixel CCD amplifier segment; a bright image of an 80 pixel diameter circular spot (\textit{left}) exposed to around 4 times full well saturation, and the average first exposure taken after the bright exposure (\textit{right}), in this case a 15s dark. The side of the segment which houses the serial register is labeled for clarity. In the \textit{left} image, the direction of the parallel readout transfer and the serial transfer are shown with arrows. Important residual image features have been labeled with numbers. 1: The bright spot imaged. The larger halo is caused by out of focus reflections in our system, although these are much too dim to create residual charge in subsequent images. 2: A bright trail read out after the spot in the same image. This trail is caused by charge leaking back into those pixels as they pass through the spot region as they are parallel shifted during readout. 3: The residual image seen in the spot region in subsequent images. As the bright spot was imaged, charge is trapped at the CCD oxide surface. This charge then leaks back into the pixels during subsequent integration causing this residual image. 4: The residual charge trail seen in subsequent images following a bright image. Charge here is trapped at the surface during the parallel transfer.
}
    \label{fig:residual}
\end{figure}

In LSSTCam CCDs there are notable residual charge images in CCDs manufactured by e2v (now Teledyne-e2v)\cite{Doherty2014}. Residual images have not been observed to the same degree in CCDs manufactured by Imaging Technology Laboratory (ITL) although there are ongoing studies into separate but similar effects due to photoresist impurities. The ITL CCDs are also operated at lower nominal parallel clock levels than e2vs which may account for some of this difference in residual charge\cite{utsumi2024lsst}. We will show that there are multiple mechanisms that enable residual charge between images in LSSTCam: one that allows charge to be trapped at the same pixels as the imaged bright object, and one that allows charge to be trapped when shifting the charge packets downward toward the serial register during the parallel transfer. These result in distinct manifestations of residual charge images as shown in Figure \ref{fig:residual}. LSSTCam CCDs all have sixteen separate segments each with their own readout amplifier. The images shown in this study are all images from single amplifier segments, not entire CCDs, as the effects discussed are all confined within a given segment. In addressing these residual charge sources we explore the parameter space of the parallel clocking scheme, how it affects these residual images. We discuss the associated downsides which include excessive leakage current when parallel clock lower voltage is brought too low (less than -6.0 V) and changes to the dynamic range that correspond to changes in the full well level when the parallel clock swing, that is the voltage difference between the high collecting and low barrier voltages, is decreased.

In the main survey, we will always know the precise location of bright stars that will cause residual charge effects. This means that residual images in the central spot region in the following images (labeled `3' in Figure \ref{fig:residual}) will always be in a known, well defined place. The effect can then either be removed when processing the images using calibrated data for that sensor, or the spot region can be removed from the survey entirely. The latter option is fairly draconian, although it will only affect saturating stars which are brighter than 14.7, 15.7, 15.8, 15.8, 15.3 and 13.9 magnitude in the $u, g, r, i, z,$ and $y$ bands respectively\cite{abell2009lsst}. These should number somewhere in the order of 100 million total in the sky\cite{luri2018gaia}, or half that in the southern sky that LSST will map. A point source will be imaged over $\sim$10 pixels on the detector. Assuming these point sources bloom on average to 10 times their point source size and that they are statistically evenly distributed over the ~40,000 square degrees of sky (which is not a perfect assumption due to the shape of our galaxy); on average we will see 25,000 of them in each LSSTCam image which will cover around 2.5 million pixels. This is only 0.08\% of the 3.2 Gigapixel camera and 8 saturating stars per amplifier segment image. So even in the case where we throw out the residual image pixels due to stars entirely it will only remove a fraction of a percent of the survey. This is not to say that masking out these pixels will be a trivial task. Failure to properly remove residual charge images will pose a significant challenge particularly for differential imaging analysis. However the area of affected pixels is relatively small for the source region of stellar objects.

The trail effects on the other hand (labeled `4' in Figure \ref{fig:residual}) are a more serious issue. Because the camera will rotate and image the same locations at different angles, finding and removing the trail signature is a more complex problem. If left undressed the trail would greatly impact source extraction and photometry by injecting linear features on the scale of 10 electrons above the sky background. If it is located, simply removing the involved pixels from the same and next image in the survey is a much more drastic solution. On average a bright source (around an estimated 3 pixels wide) will affect all pixels in that amplifier segment column read out trailing after it in the same image (`2' in Figure \ref{fig:residual}. This is caused by charge leaking into the potential wells as those wells are transferred through the overexposed spot region `1'), and trailing before it in the following image (`4' in Figure \ref{fig:residual}, caused by released residual charge trapped in this region). This means we lose at least three entire columns, or 6,000 pixels per point source, not accounting for situations when the bright source columns overlap in the same amplifier segment. In total this amounts to about 5\% of the survey affected by residual image trails from bright stars. 

In addition to bright stars, there is also the growing issue of other bright objects which have less well defined locations such as low earth orbit satellites (LEOSats). There are on the order 1 million satellites currently filed for launch during the survey\cite{falle2023one}. These will start to saturate the CCD sensors starting at 4th magnitude\cite{Tyson2020}. There are already some satellites recently launched which are as bright as 0th magnitude\cite{nandakumar2023high} which will likely bloom the entirety of any of the CCDs which it crosses. Even if only a few tens of thousands of these satellites are bright enough to saturate the full well, we will still see about one of them in every image and they will cross 12 CCD sensors on average. The fact that they cross many pixels or entire sensors exacerbates the issue of residual charge images. The mitigation and removal of residual charge due to LEOSats is then an important issue that demands attention both in the central bright image region and especially in the trail effects that they will cause which will cover large swathes of the camera focal plane.

\section{METHODOLOGY}
Residual charge images were first seen in the main LSSTCam CCD images by Doherty et al. in 2014\cite{Doherty2014} in prototype CCDs and later rediscovered during integration and testing  at the SLAC National Accelerator Laboratory in the main LSST Camera during integration and testing. These were seen both as residual charge after bright flat-field images and imaged spot sources. 

We examined this effect more in depth at the UC Davis CCD Laboratory which is equipped with the LSST f/1.2 Beam Simulator\cite{Tyson2014} which mimics the optical beam pattern of the Rubin Observatory Simonyi Survey Telescope and images photographic reticle masks that mimic astronomical objects onto the CCD. A science grade LSSTCam e2v CCD250 is installed with the same readout electronics used in the main camera\cite{snyder2024electro}. This single CCD setup allows us to perform in depth studies of sensor effects and and their dependence on various configurable properties that would not be possible on the main camera. 

A standard way to diagnose residual charge images in CCD cameras is to use a flat-field to uniformly overexpose the CCD to a signal level greater than saturation\cite{Janesick2001}. Then the residual charge can be measured in following dark exposure images as a uniform offset from the nominal readout. This method is a great way to determine the residual charge in the region where an object is imaged, but it would fail to pick up effects like the residual charge trail we see in LSSTCam e2v CCDs. This trail effect is much more detrimental to the science goals, and so we developed a different protocol to measure residual charge in both the area where the bright source was, and in the trail region.

We developed a series of tests that let us characterize the residual charge effect in all regions. We use an image mask of a single spot that is around 80 pixels in diameter. We take a series of 20 bias images and then expose the sensor to a bright spot source around 4 times the full well saturation with nominal settings in a 15s exposure. The exact level of full well is variable when parameters like voltages or readout clocking are changed. In each test we keep the light source at the same intensity, but depending on the readout settings, the relative full well level changes. In this way we are comparing the response to the same brightness of source object. This will be the best indicator of how the camera will respond to the same on-sky sources for a given change in configurable parameters. We then read out that spot image and immediately follow it with twenty 15 second dark images. We can then look at each subsequent dark to see trends in residual images. We refer to this as our ``standard dataset". The sensor at UC Davis has relatively small residual charge compared to some sensors on the main camera. Therefore we repeat this sequence of biases, bright spot image, and dark images 50 times in order to drive up our signal to noise. We keep exposure times at 15s to mimic the cadence of the LSST survey. This means that residual charge images and their scaling to specific parameters will act like those seen in the main survey. It also gives sufficient time resolution to examine the decay of residual charge, even if a longer 30s  cadence is adopted for LSST.

The specific procedure for one of these datasets is to subtract the median overscan region pixel value of each row from every pixel in that row, for every given image. We then take the mean value of each pixel in the relevant residual charge region over the 50 tests. We then subtract the overscan subtracted bias image of that area from the image. Then to calculate the residual charge in that area we take the median pixel value in that area of the coadded image. This eliminates any potential outlier pixels via the median operation while still giving sub-electron resolution with the mean operation. Our errors are the standard deviation of the mean of pixel values in the given region. Due to bias instability we also scale the residual by the median value measured for the last half of the 20 coadded darks. This gives all of our tests the same baseline on a long enough time scale. This increases the error in quadrature by the standard deviation of residual measured in those last darks. The two main regions of interest we sample are the central spot region where the bright spot was imaged, and the residual charge trail region between the spot and the serial register. These are labeled `3', and `4' respectively in Figure \ref{fig:residual}. 

We also report the median maximum readout we achieved in the coadded bright spot area (labeled `1' in Figure \ref{fig:residual}) of the oversaturated spot image. This is a stand-in for our full well pixel value which is a measure of our system's dynamic range which may be impacted by mitigating residual charge. This is not the standard way of calculating the full well value of a sensor which usually entails finding where the photon transfer curve falls off from linear scaling as a function of source brightness. However our method should be a close analog for how the full well value scales in our parameter space and, importantly, it is easily measurable using only our standard dataset. This method is limited by potential variations in gain between different sets of operating parameters.

The parameter space we explore is configuring the clocking voltages and timings of the CCD readout. Specifically we explore the parallel clocking voltage levels which are directly related to residual charge\cite{utsumi2024lsst}. It should be noted here that LSST CCDs are fabricated on 100$\mu$ thick p-type silicon with 5-10k$\Omega$-cm resistivity. A substrate bias voltage, applied from the front side of the device, biases the thin conductive window on the back side and results in a fully-depleted substrate throughout the imaging area with electric field of 5–10 kV/cm. This drifts photoelectrons toward the readout on the front side. In all tests in this study and under standard operating parameters, LSSTCam is operated with a constant bias voltage of -50 V. We tested many clocking voltage schemes in our study. Most of these were formatted by the LSST Camera team to ensure proper protocols were followed to prevent readout failure. The voltages that are relevant to an understanding of our work are mainly the three listed in Table \ref{tab:voltages}. One important voltage we use is the P-Up configuration which uses the former nominal voltage configuration, but with both parallel voltages shifted higher. This allows us to exacerbate the residual charge image effects in our sensor which normally shows relatively small residual charge compared to some LSSTCam CCDs. In this way we greatly increase the signal to noise in our measurements. We also list the values used in the nominal settings used before this work and the final new configuration which has been adopted for for use in the main survey. 

\begin{table}[h]
    \centering
    \caption{Relevant voltage configurations used in probing residual charge in LSSTCam e2v CCDs. The nominal voltages are the ones used by the main camera before this work. The new configuration is likely the final one used in the main camera for the survey which was developed by the LSST Camera team to eliminate residual images by the mechanism described in this study. All but P-Up follow Rubin observatory guidelines to prevent readout failure. In those, we keep the parallel barrier voltage (Parallel Low) at -6.0 V which is the lowest the main observatory is willing to go. P-Up is the nominal configuration with the parallel voltages both raised in order to exacerbate the residual charge effect.}
    \vspace{0.5cm}
    \begin{tabular}{|c|c|c|c|}
    \hline
         Clock Name & Nominal Voltages & New Voltages & P-Up Voltages\\ \hline \hline
         Parallel High & 3.3 V & 2.0 V & 5.0 V\\
         Parallel Low & -6.0 V & -6.0 V & -4.2 V\\
         Serial High & 3.9 V & 3.55 V & 3.9 V\\
         Serial Low & -5.4 V & -5.75 V & -5.4 V\\
         Output Gate & -3.4 V & -3.75 V & -3.4 V\\
         Output Drain & 23.4 V & 22.3 V & 23.4 V\\
         Reset Drain & 11.6 V & 10.5 V & 11.6 V\\
         Reset Gate High & 6.1 V & 5.0 V & 6.1 V\\
         Reset Gate Low & -4.0 V & -5.0 V & -4.0 V\\
         \hline
         
    \end{tabular}
    \label{tab:voltages}
\end{table}

We also do a number of tests where we keep almost all the voltages constant at the P-Up or nominal values, but vary only the parallel high, or parallel low voltages to find their specific impact on residual charge. As shown in Table \ref{tab:voltagesweep},  we change the parallel voltages individually to quickly find direct voltage dependencies without small secondary effects from other changes. These tests are useful in determining trends but they are not potential configurations for final camera operations. 

\begin{table}[h]
    \centering
    \caption{In order to measure the dependence of residual charge on parallel voltages, we leave all voltages at the nominal values and then independently change the parallel high collecting voltage to the values in the first row. We then independently sweep through different parallel low values in the second row with all other values at the nominal. The third row is a voltage sweep where we start with the P-Up configuration and lower the collecting voltage.}
    \vspace{0.5cm}
    \begin{tabular}{|c||c c c c c c c|}
        \hline
        Parallel High Sweep & 3.0 V & 3.2 V & 3.4 V & 3.6 V & 3.8 V & 4.0 V & 4.2 V\\ \hline
        Parallel Low Sweep & -5.8 V & -6.0 V & -6.2 V & -6.4 V & -6.6 V & -6.8 V & -7.0 V\\ \hline
        P-Up Parallel High Sweep & 3.8 V & 4.0 V & 4.2 V & 4.4 V & 4.6 V & 4.8 V & 5.0 V \\ \hline

    \end{tabular}
    \label{tab:voltagesweep}
\end{table}

\begin{figure}[h]
    \centering
    \includegraphics[width=0.8\linewidth]{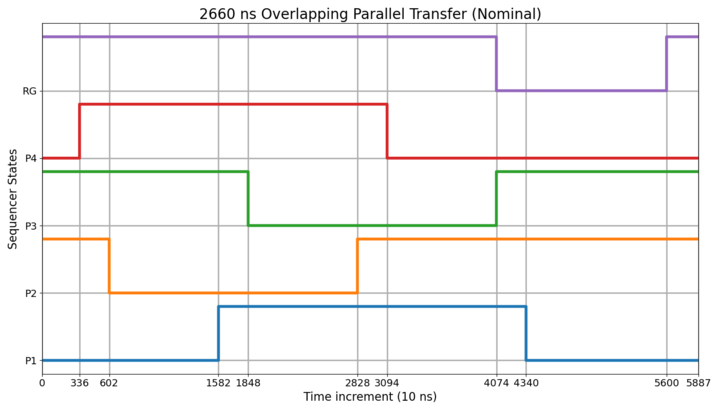}
    \caption{The clocking scheme for the four clocking gates (P1, P2, P3, P4) and the reset gate (RG) during parallel transfer and readout in LSSTCam e2v CCDs. When the gates change phase, three gates are held at the high collecting voltage for a small 2660ns period. This may cause charge to reach the CCD surface during the parallel transfer. This overlap timing is a configurable parameter we can alter to measure its influence on the residual charge trail.}
    \label{fig:Pclks}
\end{figure}

We can also alter the timing of the CCD readout, and the parallel transfer in particular. In the nominal LSSTCam e2v CCD clocking scheme, when each of the four parallel clocking gates change phase from the low barrier phase to the high collecting phase, there is a short 2660ns period where three of the four gates are held at the collecting voltage at the same time before the next gate changes to the low barrier phase. The gate overlap period accounts for about 23\% of the total transfer timing with the nominal readout timing. This standard parallel clocking scheme is shown in Figure \ref{fig:Pclks}. This scheme gives the charge ample time to transfer, but may be related to the novel trail region residual charge. We do not know the exact mechanism that would cause this but we have a few hypotheses. One possible cause of this would be that, as the three clocks are held high, charge interacts more with the CCD surface. This would cause residual images in the trail region. If this is the case, it is a subtly different mechanism than the standard cause of residual charge images in the spot region during integration. We change the clock timing to decrease the duration of this parallel clock overlap and measure its effect. 

\section{RESULTS}

\begin{figure}[h]
    \centering
    \includegraphics[width=0.7\linewidth]{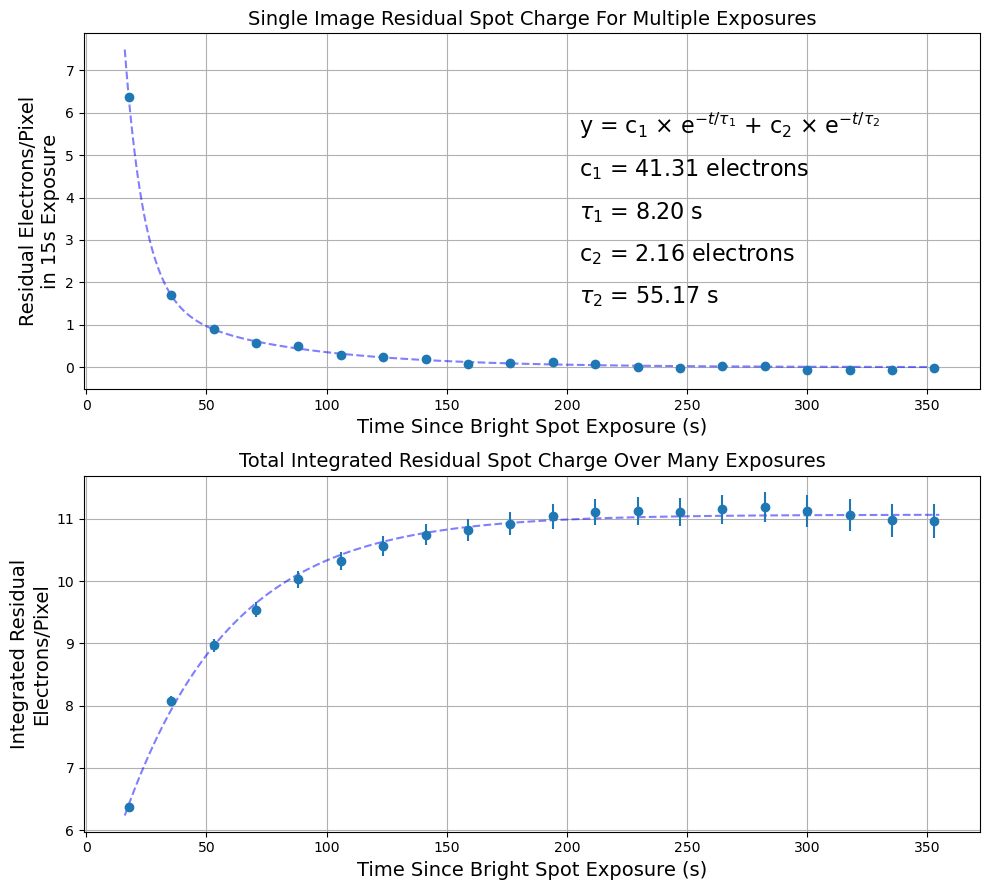}
    \caption{Residual charge as a function of time in two frames. \textit{Top} shows the single image frame which measures the charge in each individual dark following a bright spot exposure. \textit{Bottom} is the total charge frame which shows the integrated charge seen in the dark images up to that point. Both are useful frames in which to view the exponential thermal decay of electrons from the CCD surface into the pixels. We will mainly use the single image frame in this work unless otherwise noted. This figure shows the residual charge in the spot region. The trail region residual charge scales similarly. Residual charge has been seen in up to the first ten 15s images and possibly more at a lower level. Changing voltage levels and clocking schemes changes the magnitudes of the charge, but the decay times largely remain unchanged.}
    \label{fig:frames}
\end{figure}

There are two useful frames in which to view residual charge. The first is the ``single image frame" which is the measured residual charge in a given dark image. Unless otherwise noted, our reported values use this frame and report the value in the first 15s exposure following the bright triggering image. This frame is useful because it is the closest comparison to what we will see in actual LSST on-sky data for a given configuration. The other frame is the ``total charge frame" where we instead add all the charges seen in images up to that time. This frame is useful for visualization purposes and for calculating thermal decay time scales for electrons returning from the surface in the silicon. A plot of residual charge as a function of time for both frames is shown in Figure \ref{fig:frames}. In this Figure we fit the residual charge with a sum of two exponential decays.
There are a few possible reasons our data is better fit by these two exponentials rather than one. Chief among them is that there may be multiple electron trap impurities at the CCD surface. These can have different characteristic decay times which means that decay from different trap impurities dominate on different time scales. This seems to hold for residual charge in both the spot and trail region. There may also be variations in trap density along the surface. 
It is important to note that it can take well over a hundred seconds (around ten 15s exposures) for the residual charge to completely dissipate from the surface, or at least to a level we can no longer detect. This decay rate, unlike the amount of charge trapped, is dependent on the kinetics of silicon at a given temperature and is not affected by changes made to the clocking timing or voltages.

\begin{figure}[h]
    \centering
    \includegraphics[width=0.7\linewidth]{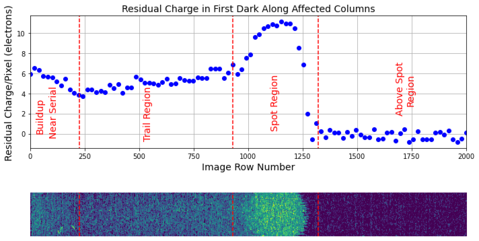}
    \caption{A visualization of residual charge per pixel as a function of row number in affected columns of a given amplifier segment exposed to a large bright spot at around four times saturation. The serial register is to the left just below row number 0. The lower image shows the actual image of the affected columns. Four regions are shown separated by vertical red dotted lines. These are, from right to left: (1) the region above the spot where we do not see residual charge, (2) the spot region where the bright spot was located, (3) the residual charge trail as it changes across rows, and (4) the end of the residual trail where residual charge appears to build up right before the serial register. }
    \label{fig:columns}
\end{figure}

We can also examine the structure of the residual images across the affected rows and columns. Taking only the columns the spot region had been saturated and averaging together the pixels in a given row, we can create a measure of residual charge across each affected row to produce Figure \ref{fig:columns}. This plot specifically uses data collected with the parallel clock overlap set to around half of its nominal value which just happened to be a dataset where all features and trends in this visualization are accentuated. The same trends exist at other values but are not always as prominent.

One can see that if we look along the affected columns, we can visibly identify four distinct regions which behave differently in Figure \ref{fig:columns}: 

(1) Above (to the right of) the spot we see that the residual is not present. In actuality, taking enough data has revealed that there is a very slight, non-zero, residual charge in this region in the first image following a bright source, although it is at a sub-electron level, even with this accentuated P-Up voltage levels. We believe this residual is consistent with charge leaking into those pixel regions as they are clocked through the spot and trail regions in the time it takes to read out the CCD.  

(2) Next there is the spot region. The dotted lines identifying this region in Figure \ref{fig:columns} are assigned by finding the highest and lowest row where the source reaches saturation level in the bright spot image. Towards the top and bottom of this region, there are bloomed pixels which reach the blooming full well condition but they are not actively being exposed to the bright source. This selected region may then extend slightly beyond the true residual spot image if those pixels are not exposed to enough light to produce residual charge. The algorithm to find the spot region then includes this exterior region where bright spot pixels reached saturation but there is no residual image seen. The spot region results therefore may be biased towards these low values. This is not true of the trail region (see below). This means that results for residual charge in the spot region are comparable to other reported spot region values, but are not comparable to trail region values.

(3) The trail region in reported values is selected by taking a window of all the rows below the spot region with a small, 10 pixel buffer on the edges and top to prevent biasing the data with the edges which may not be affected to the same degree. In data analysis, typically both the regions labeled ``Trail Region" and ``Buildup Near Serial" are combined when discussing the residual trail and readout values. They are separated here to show that there is a general fall off in residual charge as we move from the spot to the serial register until we reach the buildup structure. This makes sense for a trail mechanism where charge is trapped when clocking the original spot image through these pixels. As the spot is clocked towards the serial region, some charge is trapped which decreases the overall charge in the spot pixel charge packets. This in turn decreases the force on the charge towards the surface and moves the charge packet farther from the surface. This results in less trapped charge at the lower rows. This structure would have to be studied for each amplifier in the camera if the camera team intends to reduce this systematic out of the main dataset. Results later in this work may show that this is not necessary.

(4) The region of buildup near the serial region is treated as structure within the trail. The cause of this structure is not well understood. It is a low level effect that is not always detectable in all voltage configurations. It is likely related to an inefficiency in transferring charge to the serial region in the initial spot image, perhaps forcing more charge to the surface in this region. It is possible this effect has been eliminated by other changes made to the clocking scheme which address some issues with clearing the serial register. This should be verified in future tests.

\subsection{Parallel Clocking and Residual Charge}
\begin{figure}[h]
    \centering
    \includegraphics[width=0.45\linewidth]{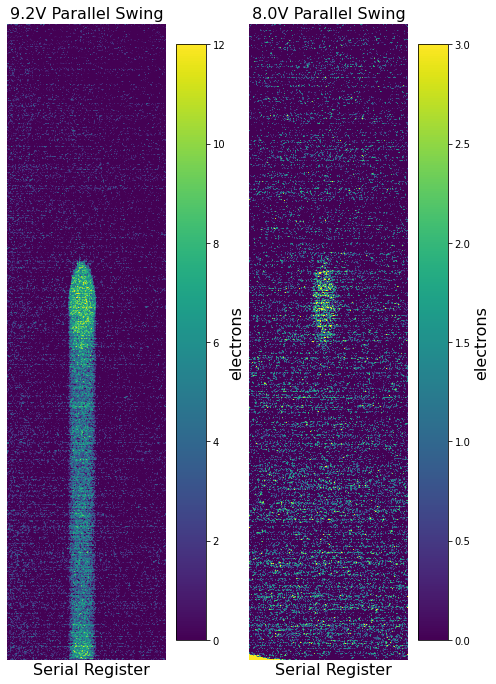}
    \caption{A comparison of the residual charge image seen in the first 15s dark following a bright spot when we change the parallel swing by decreasing the high collecting voltage. Here we change from a 9.3V parallel clock swing on the left (which was the nominal value for LSSTCam when undergoing testing at SLAC) to an 8.0V parallel clock swing on the right, which is the final configuration settled upon largely due to these results. It should be noted that the plots have different colorbar scales which allows the spot to be seen in both but also makes the noise in the right image appear larger even though it is not. Here we use the P-Up voltage scheme which greatly exacerbates the residual image effect. Performing the same decrease starting at lower voltages like the nominal shows a similar scaling of residual charge. Decreasing the parallel swing in this way virtually eliminates the residual charge trail, even with P-Up settings, and it greatly decreases the median residual charge in the spot region.}
    \label{fig:swingimage}
\end{figure}

Now that we have an understanding of residual charge in both space and time within the CCD, we discuss the actual results of varying our previously discussed alterable parameters. We first examine the dependence of residual charge images on the collecting and barrier parallel voltages. As would be expected, the general trend is that decreasing the parallel high collecting voltage also decreases the amount of residual charge in images following a bright spot in both the spot and trail regions. This is because decreasing the collecting voltage decreases the force pulling charge to the surface. Decreasing the parallel swing also decreases the residual charge seen. This is because the full well level is also decreased which allows more charge to bloom before reaching the surface. This effect is shown as a comparison of CCD images when changing the parallel swing via lowering the high collecting voltage in Figure \ref{fig:swingimage}. This effect holds whether it is the barrier or collecting voltage that is changed, although the effect is of course greater when changing the collecting voltage.

\begin{figure}[h]
    \centering
    \includegraphics[width=0.7\linewidth]{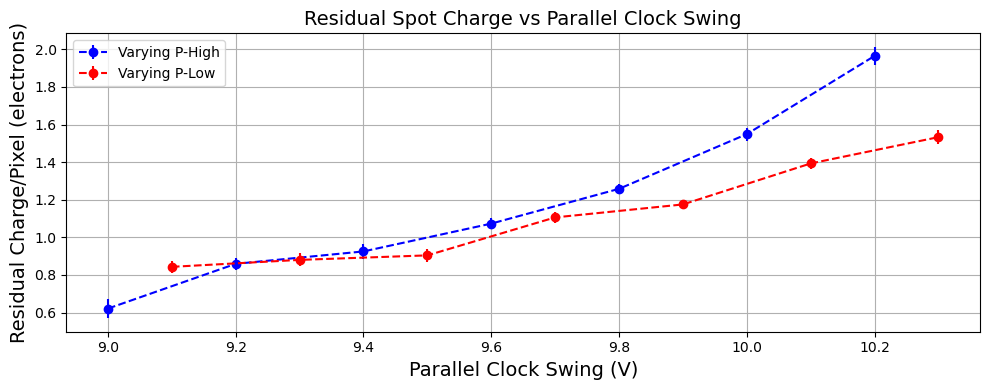}
    \caption{A comparison of how residual charge scales with parallel clock swing when varying both collecting (P-High) and barrier (P-Low) voltages independently. In both voltage sweeps, all other voltages are held at the nominal values. Varying the barrier voltage changes the residual largely based on the number of electrons that can be collected in a given pixel. Varying the collecting voltage has a larger impact on residual charge because it additionally pulls the charge packet closer to the CCD surface as it is increased.}
    \label{fig:PHvsPL}
\end{figure}

As a direct comparison of how the residual charge scales when we change the parallel high collecting voltage compared to the parallel low barrier voltage, we use the data collected from the two voltage sweeps in Table \ref{tab:voltagesweep} where we vary the collecting and barrier voltages independently and keep all others at the nominal. As shown in Figure \ref{fig:PHvsPL}, varying the collecting voltage has a much more pronounced effect on the measured residual charge. This is because while changing either alters the swing, changing the collecting voltage also directly alters how close the electron charge packet in a given pixel is pulled towards the surface.

\begin{figure}[h]
    \centering
    \includegraphics[width=0.7\linewidth]{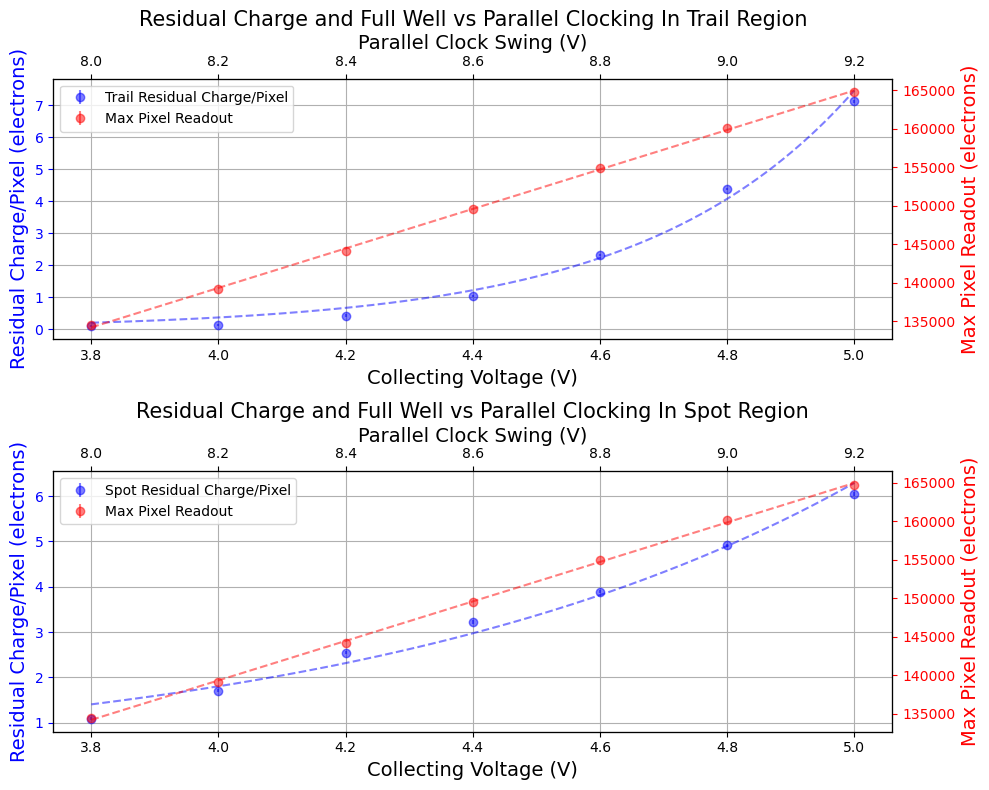}
    \caption{Residual charge in the first dark following a bright spot image, and the max pixel readout in the initial bright spot as we vary the high collecting voltage in the parallel clocking from the P-Up voltage configuration. The latter is a stand in for full well value. Decreasing the full well decreases the camera's dynamic range which is a big trade-off in this method of reducing residual charge. The \textit{top} plot is the residual charge in the trail region and the \textit{bottom} is in the spot region. In both plots there are two vertical axes, the \textit{left} which shows the units of the residual charge measurement and the \textit{right} which shows the units of max pixel readout. There are two horizontal axes, the bottom which shows the collecting voltage value, and the top which shows parallel clock swing. We fit a line to the max pixel readout values and an exponential to both the residual charge scalings to show that the residual charge falls of much quicker than the full well, especially in the more concerning trail region. Reported trail and spot values are measured differently and therefore not directly comparable.}
    \label{fig:fullwellPH}
\end{figure}

Decreasing the collecting voltage in the parallel transfer is one of the most effective ways to decrease the residual charge effect seen in e2v CCDs in LSSTCam. But it also has the drawback that it lowers the amount of charge able to be collected in a given pixel. This lowering of the full well level means that the detector's dynamic range is decreased at the bright end. We expect this effect to be fairly linear with the parallel clock voltage swing as that directly determines the pixel well size. In Figure \ref{fig:fullwellPH} we show the results of collecting data from the P-Up parallel high sweep from Table \ref{tab:voltagesweep}. We vary the high collecting voltage around the P-Up voltage configuration and measure trail and spot residual charge in the first dark image. We also measure the average maximum pixel readout value in the saturated spot in the initial bright image and use this average maximum pixel value as a stand-in for full well value. We are using values near P-Up setting here for larger signal to noise and to emulate CCDs with worse residuals. Similar studies were performed starting with lower parallel collecting voltage levels, but there is not much to be gained by plotting the residual charge levels from these since the trends are largely the same and the noise level is much larger.

Although the dynamic range of LSSTCam will be decreased by lowering the parallel clock swing, the rate at which the maximum pixel readout falls when decreasing the collecting voltage is linear, whereas the residual charge measured decreases exponentially. This is especially pronounced, as evident in Figure \ref{fig:fullwellPH}, in the trail region which is the greater threat to the LSST's systematic limits. This means that while there is a trade-off to decreasing the parallel high collecting voltage, there is proportionally substantial value gained. Any decrease in the collecting voltage will exponentially decrease the systematics in the survey caused by residual charge images, especially in the trail region. The loss in dynamic range will also be at the bright end. This is important because the main mission of the Rubin observatory is not to get better images of known bright objects, but rather to have a wide survey of dimmer objects and fainter transient objects than have been seen in previous surveys. In presenting this work to the LSST camera team, one of the outcomes was that the camera voltage scheme was changed from the nominal 9.3 V parallel swing to a lower 8.0 V parallel swing by decreasing the high collecting voltage. Other voltages were changed in conjunction to arrive at the configuration called ``new" in Table \ref{tab:voltages}. This change resulted in the virtual elimination of residual charge images in the trail region and a large decrease in the spot region. A thorough study still needs to be done using the main LSSTCam sensors to determine the exact quantitative outcomes. We expect the residual charge to scale similarly to that in our system which demonstrated around an 80\% decrease in the spot region. This is likely sensor dependent and our lab only has access to one e2v CCD.

\begin{figure}[h]
    \centering
    \includegraphics[width=0.7\linewidth]{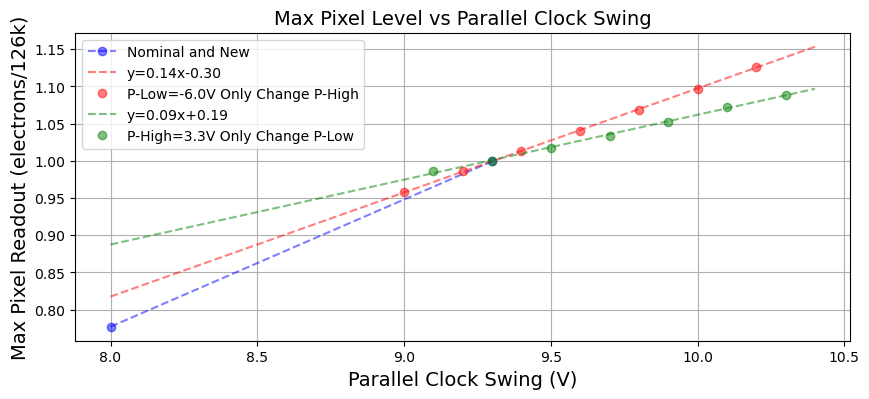}
    \caption{Average maximum pixel readout in the bright spot region as a function of parallel clock swing in a few different sets of tests. This is a stand in for the impact on the dynamic range due to changing the parallel clocking voltages. Here the vertical axis is normalized to the value for the nominal voltage configuration results. The red and green datasets are made by changing only the collecting (P-High) and barrier (P-Low) voltages respectively while keeping all other voltages at constant nominal values. Decreasing the parallel voltage sweep via lowering the P-High voltage corresponds to a smaller decrease in the dynamic range than increasing P-Low by the same amount. We show the nominal and new voltage configurations in blue. The maximum pixel readout level is decreased to $\sim$80\% of the nominal in the new voltage scheme although this varies slightly between amplifier segments and would also vary between CCDs in LSSTCam.}
    \label{fig:fullwellHvL}
\end{figure}

A final notable feature of the impact of varying the collecting vs barrier voltages is their different impact on the full well level. Decreasing the swing by changing either voltage decreases the full well and dynamic range of our camera. However, as shown in Figure \ref{fig:fullwellHvL}, increasing the low barrier voltage more effectively decreases the dynamic range than lowering the collecting voltage. The importance of this is that, just as it is more beneficial to mitigating residual charge, lowering the parallel high collecting voltage is less negatively impactful on the full well level. This also means that lowering both voltages while keeping the swing constant will actually benefit the dynamic range while simultaneously decreasing the systematic effects of residual charge. The reason we only go as low as a barrier voltage of -6.0 V is because of concerns of leakage current if we go lower. We were advised by the CCD manufacturer, e2v, that increased current flow due to operating the CCD at a lower barrier voltage could result in damage or breakdown of the sensor. In contrast, the ITL LSSTCam CCDs are operated at a barrier voltage of -8.0 V and higher current flow has been measured for those sensors although they experience virtually no residual surface charge effects. In our system at UC Davis, we successfully went as low as a barrier voltage of -7.0 V with promising results although a exhaustive testing protocol on other systematics was not done for that configuration.

 \subsection{Parallel Clock Overlap Time and Residual Charge}

Other than changing parallel clocking voltages, the main configurable parameter and independent variable we have to explore is the amount of time that three parallel clocks are held high when gates change phase. We refer to this as the parallel clock overlap. We take our standard dataset with our clocking sequence changed to keep the overall timing to shift down one pixel the same. We change the overlap by varying amounts from nominal to no overlap. It should be noted that it takes time for the clocks to switch phases and this is not included in the considered overlap time. However this time is on the order of 5-10 ns and should not significantly impact our findings. It also takes a comparable time to switch phase in either direction, so the overlap due to this ramp up timing is very limited.

\begin{figure}[h]
    \centering
    \includegraphics[width=0.45\linewidth]{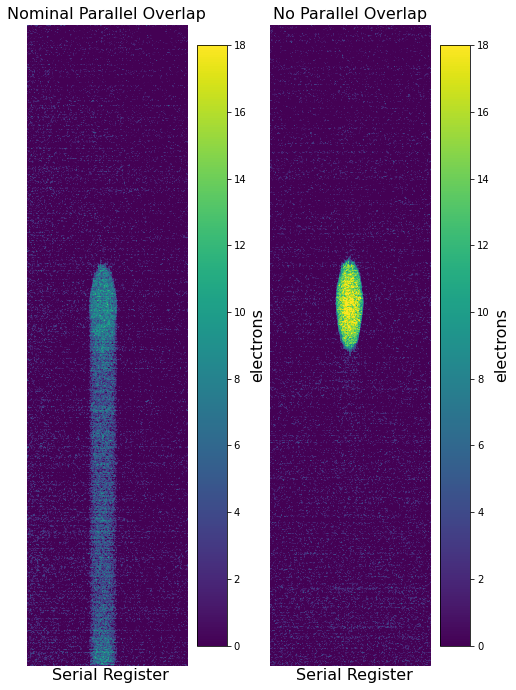}
    \caption{A comparison of the residual charge image seen in the first 15s dark following a bright spot when we change the parallel transfer clocking timing to eliminate the overlap time when three gates are held high during phase changes. Decreasing or removing the overlap time increases the residual charge seen in the spot region but decreases that seen in the trail region. This points to two separate mechanisms which cause residual charge in these regions which is related to the parallel transfer clocking.}
    \label{fig:overlapimage}
\end{figure}

Altering the parallel clock overlap time does have a strong effect on the residual charge seen in images. However it is not as straightforward as in the case of altering the parallel clocking voltages. Changing the overlap timing affects the residual charge in the spot and trail regions differently. In the spot region, decreasing the parallel overlap increases the residual charge measured, whereas in the trail region, the residual charge decreases as the overlap is decreased. In Figure \ref{fig:overlapimage} we show these results which point to separate mechanisms that drive residual charge in the spot and trail region.

\begin{figure}[h]
    \centering
    \includegraphics[width=0.7\linewidth]{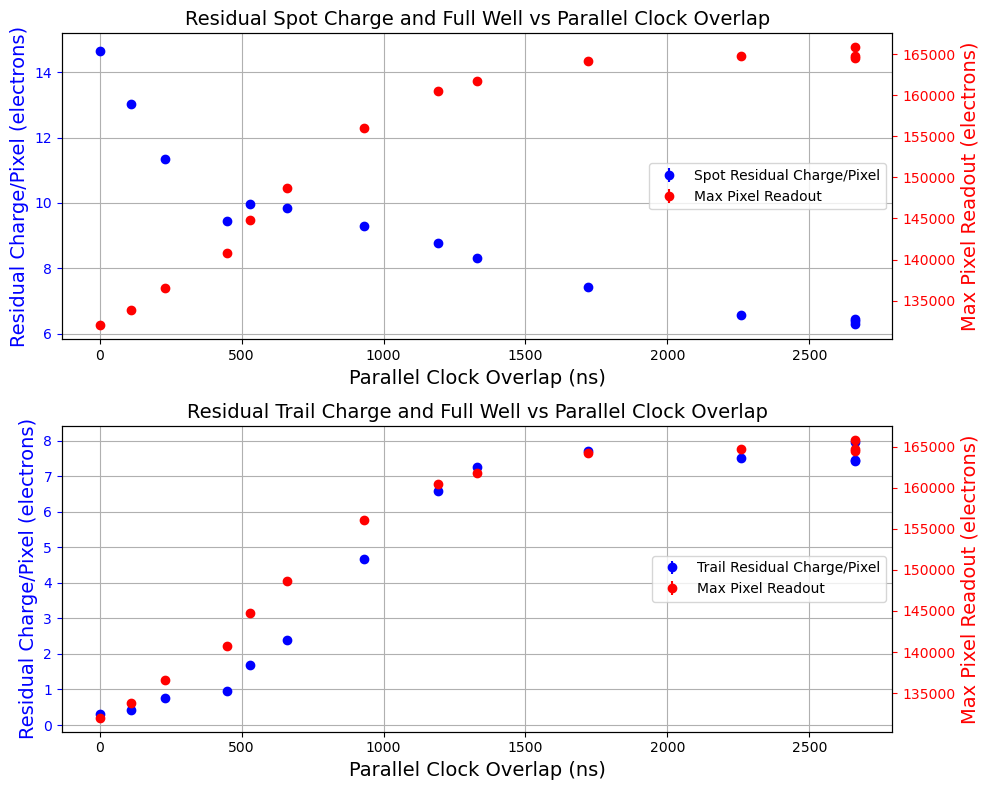}
    \caption{Residual charge and maximum pixel readout as functions of parallel clock timing overlap. The top chart shows the residual charge in the spot region and the bottom shows the trail region. While spot residual charge increased by decreasing the parallel clock overlap, the trail residual charge decreases with some apparent plateaus at the high and low ends. This may indicate that the effect takes some time during the overlap period to reach full impact. }
    \label{fig:overlapvresidual}
\end{figure}

We performed several tests at varying levels of parallel clock timing overlap. Both the spot and trail region have strong dependencies of this value but with very different forms. As shown in Figure \ref{fig:overlapvresidual}, the trail residual charge decreases with the overlap time while the spot residual increases. But while the spot residual seems to follow an somewhat exponential pattern, the trail residual scales exhibits fairly logistic growth. In the trail's case we believe this is because whatever effect is causing the charge to be captured has some amount of overlap time required to fully take effect. After that time is provided the residual charge level plateaus. 

The reason why the parallel clock timing overlap effects the residual charge images is not perfectly understood. We explored this parameter mainly because the trail effect clearly occurs during the parallel transfer. One hypothesis is that holding three gates high increases the realized potential in the silicon and pulls the charge packet closer to the CCD surface during the transfer. This causes charge to be captured during readout in the trail region. However, preliminary simulations of the charge packet position within the CCD using the Poisson\_CCD package\cite{lage2021poisson_ccd} show that charge does not move closer to the CCD surface when three parallel gates are held high. In fact, the charge packet moves further from the CCD surface as there is more area for the electrons to spread out. 

One possibility is that this spreading out of charge may allow charges in the readout trail to interact more frequently with a spatially rare (on the scale of the saturated PSF) dilute family of traps along the surface. This would allow charge to interact with these rare traps over a wider area even if the charge packet as a whole is further from the surface.

We also operated the CCD in ``scan mode"\cite{juramy2014driving, o2016integrated}, which allows us to read out the raw video pulse for each pixel read into our readout electronics. We found that there is no significant change in the signal shape when we change or eliminate the parallel overlap. This indicates that it is not an issue of giving the charge packed adequate time to transfer. It was also considered that the additional residual charge in the trail region is caused by additional time that the charge packet spends in a given region of silicon. This is certainly not the case since the increase in time due to parallel clock overlap is not proportionally long enough.

It is also important to note how closely the residual charge in the trail region scales with the maximum average pixel readout in the bright spot shown in Figure \ref{fig:overlapvresidual}. The full well seems to similarly reach a plateau level around the same value of parallel clock overlap which is likely due to charge having ample time and area to transfer without blooming during the transfer process. The correlation implies that the impact on residual charge in this region may be scaling with the full well value and the mechanism may simply be a decrease in charge packet size which moves the packet farther from the surface. However, this does not fully account for why we are able to see the residual charge images in the trail region in the first place when it can only be occurring during the short time of the parallel transfer. It also does not explain the scaling of the residual charge level in the spot region. In the end we may not have enough resolution to measure the correlation between the full well plateau and the residual charge plateau due to changes in parallel clocking overlap as we are operating the CCD in a region where the readout is extremely nonlinear.

There are a few remaining possibilities. It is possible that holding three gates high at the same time is actually forcing charge that was captured in the spot region out from the surface. That charge then does not all reach the charge packet and some of it is pulled along closer to the surface during readout, and is captured in the surface in the trail region as the main spot charge packets are clocked towards the serial register. This would also explain why the residual charge in the spot area increases as the trail residual decreases.

Another possibility is that while the charge is not actually moving closer to the CCD surface when three gates are held high, the charge packet is oscillated towards and away from the surface when we change the number of gates held high. In the clocking scheme where only two gates are held high (no parallel overlap), this would not occur and the charge packet would remain at the same relative distance from the surface. If this is the case, that oscillation may actually cause some charge to move too far and interact with the CCD Si-SO$_2$ surface.

As far as a practical measure to remove residual charge persisting in LSSTCam images in the survey, decreasing the parallel clock overlap is not as clearly effective as altering the parallel clock voltage levels. The increase in the spot region residual charge and the uncertainty of the root mechanism mean that there are more clear drawbacks than simply decreasing the dynamic range, and there might be more unknown drawbacks if more extensive studies and sensor characterization is done with these modified settings. It is likely that if the goal is to minimize the residual charge effects, especially in the more problematic trail region, there is some ideal optimization of decreasing the parallel clock overlap in conjunction with altering the parallel transfer clock voltages. 

\section{FUTURE WORK}
As we have discussed, there are still open questions regarding the source and mechanism that drives residual charge in LSSTCam e2v CCDs. From an instrumentation standpoint this is the big remaining question which may be the subject of future projects. If the mechanism driving residual charge is found, it may lead to better understanding of camera sensors and how to mitigate this effect in readout configurations or in the manufacturing process. However, from the standpoint of the LSST Camera team, this is not the most pressing issue. LSSTCam is already built and we have characterized residual charge and how to mitigate it in survey images. But we have not eliminated it completely. 

Residual charge images will appear in the LSST. They are very likely going to be below the detection threshold for a single image and they will not be in the same place for multiple images. When LSSTCam goes on-sky it will be necessary to verify that residual charge images are not a significant systematic error. Residuals at the sub-electron level in coadded exposures may be of concern, since surface brightness systematics at this level on several arcsecond scales can affect some science. This will require a fairly in depth study of bright source images and the following images in both the spot and trail regions. This will be confounded by sources in those regions which are not necessarily known or able to be removed from biasing the dataset. The same image reduction algorithm we use in our lab study will not be usable as sky noise and dim sources will offset images from bias frames or dark frames taken before the bright source. That is not an issue in the controlled dark system at UC Davis. This sky offset may be mitigated by subtracting the average pixel value of the suspect regions by the average of a similarly sized region of the same image, which will have a similar sky level and on average the same number of sources. With a sufficiently large sample size, which the LSST will certainly have, we may be able to detect this systematic if it exists and even characterize it for removal in each individual amplifier. If necessary, the process of removal will pose a problem of characterization and subtraction while limiting residual errors. This has been dealt with in other surveys such as Pan-STARRS' burntool program which removes residual charge signatures using a one-dimensional exponential model\cite{waters2020pan}.

 \section{CONCLUSION}
 LSSTCam e2v CCDs exhibit residual charge effects in images following exposure to bright sources. These manifest not just as ghosts in the same region of the focal plane as the source, but also as a trail between the source and the serial register. This indicates charge is being captured in the CCD and likely on the CCD's oxide surface layer interface not only during integration, but also during the parallel transfer and readout. Both these regions, but especially the trail residual charge images will create a large systematic error in the Rubin Observatory Legacy Survey of Space and Time if not addressed. The trail effect in particular would inject linear contrasting features on the scale of 10 electrons above the sky background which would greatly impact source extraction and photometry.

 Through our studies into potential causes of this systematic, we developed a technique for studying residual charge images using bright spot sources rather than flat-field images as is standard for this kind of study\cite{Janesick2001}. Using this methodology we determined a few parameters which have strong correspondences with the amount of residual charge seen in images. Importantly, these parameters do not always correspond to similar behavior of residual charge in the spot region compared with the trail region. The most evident of these parameters are the levels of the voltages used for parallel transfer clocking, and the timing of the phase changes of gates during the parallel transfer. Specifically in this latter case, there is an overlap period where three gates are held at the high collecting voltage each time the gates change phase.

 The alteration of parallel transfer clocking voltages has a well understood effect on residual charge images. Decreasing the the swing between the high collecting and low barrier voltages decreases pixel well size and therefore the maximum number of electrons in a given pixel. This in turn increases the minimum distance the charge packet can get from the CCD surface and makes electrons less likely to be captured in a given time period. Decreasing the collecting voltage also increases the minimum distance between the charge packet and CCD surface. We can therefore greatly reduce or eliminate the residual charge seen in LSSTCam e2v images by decreasing the parallel transfer clock swing via lowering the high collecting voltage. This comes with the trade-off of decreasing the dynamic range of the instrument at the bright end.

 Reducing or eliminating the parallel clock overlap time also significantly reduces the residual charge trail. However this comes not just with the loss of dynamic range of the camera, but also with an increase in residual charge seen in the same pixels as the initial bright source. The exact mechanisms that causes these effects is not understood, although it does point to the parallel clock overlap being a contributing factor in how charge is captured during the parallel transfer and the resulting residual charge trail.

 Based on the results of this study it was decided by the LSST Camera team to lower the the parallel clocking voltage swing from 9.3 V to 8.0 V via decreasing the high parallel transfer collecting voltage. More extensive studies are needed to characterize residual charge in the main camera. However preliminary images show that this virtually eliminates residual charge in the trail region and greatly decreases it in the same region as the bright source. In our studies with the f/1.2 beam simulator we saw over an 80\% decrease in residual charge in that same region for the same change in voltages.

\acknowledgments

 We acknowledge significant help from Gregg Thayer, Yousuke Utsumi, Tony Johnson, Stuart Marshall and Pierre Antilogus.
 We thank Sean MacBride, Yousuke Utsumi and Andy Rasmussen for carefully reviewing our manuscript. 

 This material is based upon work supported in part by the National Science Foundation through Cooperative Agreement AST-1258333 and Cooperative Support Agreement AST-1202910 managed by the Association of Universities for Research in Astronomy (AURA), and the Department of Energy under Contract No. DE-AC02-76SF00515 with the SLAC National Accelerator Laboratory managed by Stanford University. Additional Rubin Observatory funding comes from private donations, grants to universities, and in-kind support from LSST-DA Institutional Members
 
 This research is supported by DOE grant DE-SC0009999 and NSF grant AST-2205095.

\bibliographystyle{JHEP}
\bibliography{biblio.bib}

\end{document}